\documentclass[manuscript]{aastex}

\slugcomment{Not to appear in Nonlearned J., 45.}

\shorttitle{K-Shell Photoabsorption Studies of the Carbon Isonuclear Sequence}
\shortauthors{Hasoglu et al.}

\begin{document}

\title{K-Shell Photoabsorption Studies of the Carbon Isonuclear Sequence}

\author{M. F. Hasoglu, Sh. A. Abdel-Naby, and  T. W. Gorczyca}
\affil{Department of Physics, Western Michigan University,
Kalamazoo, MI 49008-5252}

\author{J. J. Drake}
\affil{Harvard-Smithsonian Center for Astrophysics, MS-70, 60 Garden
Street, Cambridge, MA 02138}

\author{B. M. McLaughlin}
\affil{ Centre for Theoretical Atomic, Molecular and Optical Physics (CTAMOP),
                          School of Mathematics and Physics, The David Bates Building, 7 College Park,
                          Queen's University Belfast, Belfast BT7 1NN, UK}

\begin{abstract}
K-shell photoabsorption cross sections for the isonuclear \ion{C}{1} - \ion{C}{4}
ions have been computed using the R-matrix method.
Above the K-shell threshold, the present results are in good agreement with the independent-particle results of \citet{reilman}.
Below threshold, we also compute the strong $1s\rightarrow np$ absorption resonances
with the inclusion of important spectator Auger broadening effects.
For the lowest $1s\rightarrow 2p,3p$ resonances, comparisons to available
\ion{C}{2}, \ion{C}{3}, and \ion{C}{4} experimental results
show good agreement in general for
the resonance strengths  and positions, but unexplained discrepancies exist.
Our results also provide detailed information on the \ion{C}{1}
K-shell photoabsorption cross section including the strong resonance features, since very limited laboratory experimental data exist.
The resultant R-matrix cross sections are then used to model the {\em Chandra} X-ray absorption spectrum of the blazar Mkn 421.

\end{abstract}





\section{Introduction}

The inner-shell excitation and ionization features of cosmically
abundant elements fall in the spectral ranges covered by the
high-resolution X-ray spectrometers onboard the {\it Chandra} and {\it
XMM-Newton} observatories.  The detailed structure and wavelengths of
absorption resonances of a given element depend on its ionization and
chemical state.  High resolution X-ray spectroscopy of these features
can in principle be used to probe the physics and chemistry of
astrophysical plasmas. Inner-shell photoabsorption resonances have
proven particularly useful for investigating the chemical composition
of the interstellar gas in the line-of-sight toward bright sources of
X-ray continuum radiation, as demonstrated in the
pioneering study of \citet{schattenburg1986}
 \citep[see also][]{paerels2001,takei2002,vries2003,juett2004,juett2006,ueda2005,yao2006,yao2009,kaastra2009}.
 ~\citet{ness2007} were able to identify several ionization stages of oxygen in
the post-outburst circumstellar material of the recurrent nova RS~Oph
based on the prominent $1s\rightarrow 2p$ resonance.

These studies have all employed oxygen and higher $Z$ elements.
Carbon is the fourth most abundant element in the Galaxy (after H, He,
and O), but has not yet been exploited as an X-ray photoabsorption
diagnostic.  It presents a special challenge for X-ray transmission
spectroscopy.  X-ray instruments often employ visible/UV light
blocking filters based on carbon compounds that are robust to space
deployment.  These filters imprint strong C K-edge absorption
signatures on X-ray spectra, rendering difficult the disentanglement
of weaker astrophysical absorption features.  The task is hampered
further still by a current lack of data describing the expected
absorption edge structure and resonances for neutral and ionized C.

In order to investigate C absorption features,
we perform detailed R-matrix calculations of the K-shell
photoabsorption cross-sections of \ion{C}{1} - \ion{C}{4}.
These cross sections are then used to interpret the x-ray spectra
from a high-quality {\it Chandra}
observation of the blazar Mkn~421 and determine relative carbon-ion abundances.

\section{Theoretical Methodology}

The specific
processes of interest are the K-shell photoexcitation of the
carbon-ion ground state,
\begin{eqnarray}
h\nu+1s^22l^q & \rightarrow 1s2l^qnp \ ,&  \label{eq1}
\end{eqnarray}
followed by two competing decay routes.
First, there is
{\em participator} Auger decay
\begin{eqnarray}
1s2l^q np & \rightarrow &
             1s^22l^{q-1}+e^-  \ ,
     \label{eq2}
     \end{eqnarray}
where the valence electron $np$ takes part in the autoionization process;
the decay width therefore scales as $1/n^3$ and goes to zero near the K-shell threshold.
There is also the more important  {\em spectator} Auger decay
\begin{eqnarray}
1s2l^q np & \rightarrow &
             1s^22l^{q-2}np+e^-  \ ,
     \label{eq3}
     \end{eqnarray}
where the valence electron $np$ does not take part in the autoionization process, giving instead a decay width that is independent of $n$.
Spectator Auger decay is therefore the dominant decay route
as $n\rightarrow \infty$ and gives a smooth cross section as the K-shell threshold is approached; above each threshold,
K-shell photoionization to the $1s2l^q$ states occurs instead.  We note that for $n=2$, there is no distinction between  spectator and participator channels.

To account for photoionization to the participator channels, we use
the standard R-matrix  method  \citep{rmbook,rmcpc} and expand the total wavefunction in a
basis consisting of free electron orbitals $\epsilon l^\prime$ coupled to each of the ground-state and singly-excited $1s^22l^{q-1}$ target
wavefunctions, plus bound orbitals $np$ coupled to the K-shell vacancy target wavefunctions $1s2l^q$.
However, the infinite number of $1s^22l^{q-2}np+e^-$ spectator decay
channels are impossible to include implicitly in such an R-matrix expansion;
instead, they are accounted for  via an optical potential approach~\citep{augerbroad}.
The target energy for each closed channel $1s2l^qnp$
is modified within a multi-channel quantum defect theory approach as
\begin{eqnarray}
E_{1s2l^q}\rightarrow E_{1s2l^q}-i\Gamma_{1s2l^q}/2\ ,
\label{eqgam}
\end{eqnarray}
where $\Gamma_{1s2l^q}$ is the $1s2l^q\rightarrow 1s^22l^{q-2}+e^-$ Auger width.  This enhanced R-matrix method was shown to be successful in describing
experimental synchrotron measurements for \ion{Ar}{1} \citep{augerbroad}, \ion{O}{1} \citep{goro}, and \ion{Ne}{1} \citep{gorne}, and Chandra high-resolution spectroscopic
observations for oxygen ions \citep{juett2004,garcia2005} and
neon ions \citep{juett2006}.
We compute the $1s2l^q$ Auger widths by applying the Smith
time-delay method~\citep{smith} to the photoabsorption R-matrix calculation
of the neighboring $1s^22l^{q-1}$ carbon ion. Since $1s\rightarrow 2p$ photoabsorption of the $1s^22l^{q-1}$ ion gives an intermediate $1s2l^q$ resonance, the subsequent
Auger decay to the $1s^22l^{q-2}+e^-$ channel can be analyzed to
obtain the Auger width.

It is also important to obtain accurate target wavefunctions using a single
orthogonal orbital basis, and this is problematic since orbital relaxation occurs when the $1s^22l^{q-1}$ states
are excited to the $1s2l^q$ K-shell-vacancy states (the $2l$ electrons are now
only screened by one, not two, $1s$ electrons).  We account for orbital relaxation by using additional pseudoorbitals as follows.  A basis of physical
$1s$, $2s$, and $2p$ orbitals is first constructed by performing Hartree-Fock
calculations for the $1s^22l^{q-1}$ ground states.  Then multiconfiguration Hartree-Fock (MCHF) calculations~\citep{mchfbook} are performed for the $1s2l^q$ K-shell-excited states, including all configurations obtained from single and double promotions to the $n=3$ shell, to obtain $\overline{3s}$, $\overline{3p}$, and $\overline{3d}$ pseudoorbitals. All target states
are then described by a configuration-interaction (CI) expansion using all configurations
consistent with single and double promotions out of the $1s^22l^{q-1}$ and
$1s2l^q$ states, using the six $1s$, $2s$, $2p$, $\overline{3s}$, $\overline{3p}$, and $\overline{3d}$ orbitals.  Our computed R-matrix \ion{C}{2} - \ion{C}{5} target energies are shown in Tables I-IV, respectively, and compared to the currently recommended
NIST values\footnote{{\tt http://www.nist.gov/physlab/data/asd.cfm}} for the $1s^22l^{q-1}$ singly-excited states.    To our knowledge,
there are no experimental or theoretical values for the K-shell-excited
$1s2l^q$ energies for comparison except for the $1s2s^22p^2(^4P)$ state of \ion{C}{2}.  The energy of this state corresponds to the K-shell threshold
in the photoabsorption of \ion{C}{1} and was measured to be  $296.07\pm0.2$~eV by \citep{cexp}, which is in good agreement with our predicted value of 296.096 eV.

Given target wavefunctions, we construct total wavefunctions by coupling an additional electron orbital.  Standard R-matrix
calculations were then performed including only the participator channels,  and by
investigating the behavior near the $1s2l^{q-1}$ resonances, the Auger widths of Eq.~\ref{eqgam} are determined.
These R-matrix Auger widths are listed in Tables V-VIII and compared to multiconfiguration Breit-Pauli (MCBP)
and multiconfiguration Dirac-Fock (MCDF) results.  Given the Auger widths,
the spectator Auger channels are also included in the optical potential R-matrix calculations
for \ion{C}{1} - \ion{C}{4} to yield cross sections with the correct resonance widths for $n>2$.

\section{Cross Section Results}
\label{sec:CI}

\begin{figure}[hbtp]

\centering
\includegraphics[width=\textwidth]{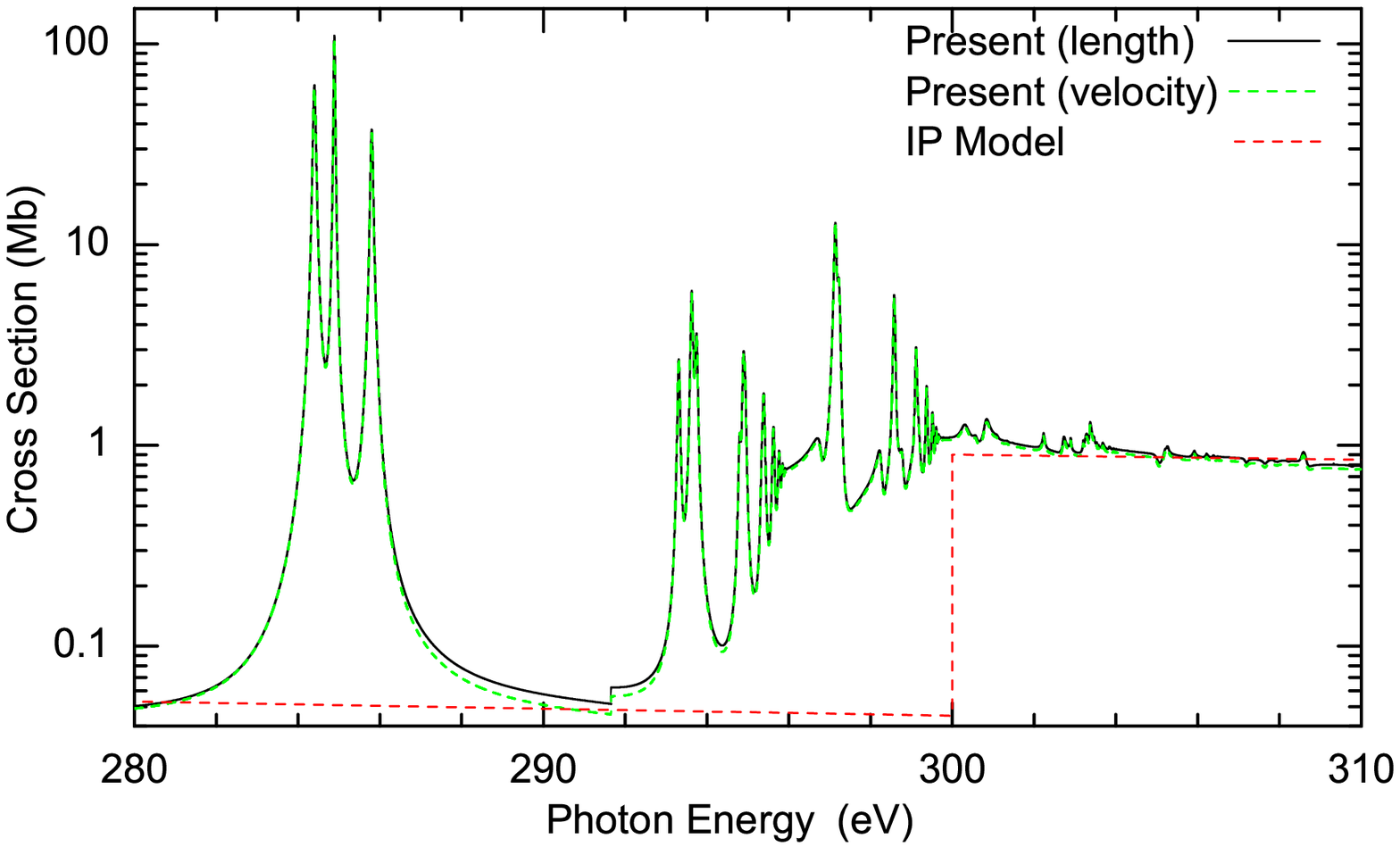}
\caption{\ion{C}{1} photoabsorption cross
sections: R-matrix length and velocity results compared to the independent-particle
results of \citet{reilman}.}  \label{cpa}
\end{figure}

Our results for the \ion{C}{1} photoabsorption cross section are shown in Fig.~\ref{cpa}.
To our knowledge, there is very limited experimental data for  comparison purposes
available in the literature \citep{janit90,janit91,henke 1993} and primarily from dual laser plasma experiments.
From their Dual Laser Plasma (DLP) experiments Jannitti and co-workers \citep{janit90,janit91} were able to make
preliminary identifications of the $1s \rightarrow 2p$  resonances in the vicinity of the K-edge.
However, it should be stressed that experimental data  on atomic carbon that does exists in the literature
 was obtained at very low resolution and performed using the Dual Laser Plasma (DLP)
 technique as pioneered by \citet{ek77}.  The dual plasma technique is useful for obtaining absorption
spectra over a wide energy range \citep{janit90,gene96}. Their
interpretation, however, can be extremely complicated due to  ions being distributed over various charge states in both
the ground and metastable-states, and the presence of a plasma can affect energy levels, as can
post-collision interactions (PCI).

Theoretical cross-sections are available in the literature from the independent-particle (IP) model \citep{reilman,yeh}
that include the direct photoionization continuum (but not the $1s\rightarrow np$ resonance contributions) as does the
scaled hydrogenic work of \citet{verner93,verner95}.
Ab initio work has also been carried out using the R-matrix approach \citep{petrini,bmcl01}.  \citet{petrini}
performed R-matrix calculations on this complex at energies above
the hole states included in their work.  No resonance structure was obtained
in their work, since they were mainly interested
in determining the shake up process (2s $\rightarrow$  2p and 3s,
and 2p $\rightarrow$ 3p excitations) in the photon energy range 25 to 45 Rydbergs,
i.e. $\sim$ 340 eV to  612 eV, which is well above the carbon K-edge.  They conclude that the 2s $\rightarrow$ 3s excitation
dominates the shake up process and is about a factor of 2 smaller than in the case
of boron \citep{badnell 1997}. In addition the 2s shake off process
(at most 20\%) would lead to the production of \ion{C}{3}  1s$^{-1}$ hole states
which can then Auger decay only to CIV 1s$^2$2p.
In contrast to this, the R-matrix work of \citet{bmcl01}  catered for resonances in the vicinity of the carbon K-edge and showed strong
resonance structure which was missing from various other theoretical approaches.

Our results for the \ion{C}{1} photoabsorption cross section, using both length and velocity
forms of the dipole operator, are shown in Fig.~\ref{cpa}.    For exact wavefunctions, the two results will be identical, and
the excellent agreement between our two cross sections is an indication that we have obtained fairly accurate
R-matrix wavefunctions.  We find similar excellent agreement between length and velocity results for
\ion{C}{2} - \ion {C}{4} photoabsorption and therefore only show length results in the remainder of the paper.
IP results are also shown in Fig.~\ref{cpa}, which are seen to be in good agreement with our R-matrix
cross section above the K-shell threshold.

Our computed $1s2s^22p^2(^4P)$ K-shell threshold at 296.10 eV is in good agreement with the
experimentally observed values of 296.2$\pm$0.5 eV by~\citep{cexp1}  and 296.07$\pm$0.2 eV by~\citep{cexp} and with the
theoretical value of  296.02 eV from a UHF-SCF  approximation.
The discontinuity around 291.6 eV in the R-matrix results is due to the turn-off of Auger broadening below $n=3$ to prevent double counting
of the Auger width of the $n=2$ resonance (the $1s2s^22p^2$ resonances only decay to participator channels).
This discontinuity of $1.1\times10^{-2}$ Mb  is very small compared to the large resonance features
of $10^1-10^2$ Mb.

\begin{figure}[!h]
\centering
\includegraphics[width=\textwidth]{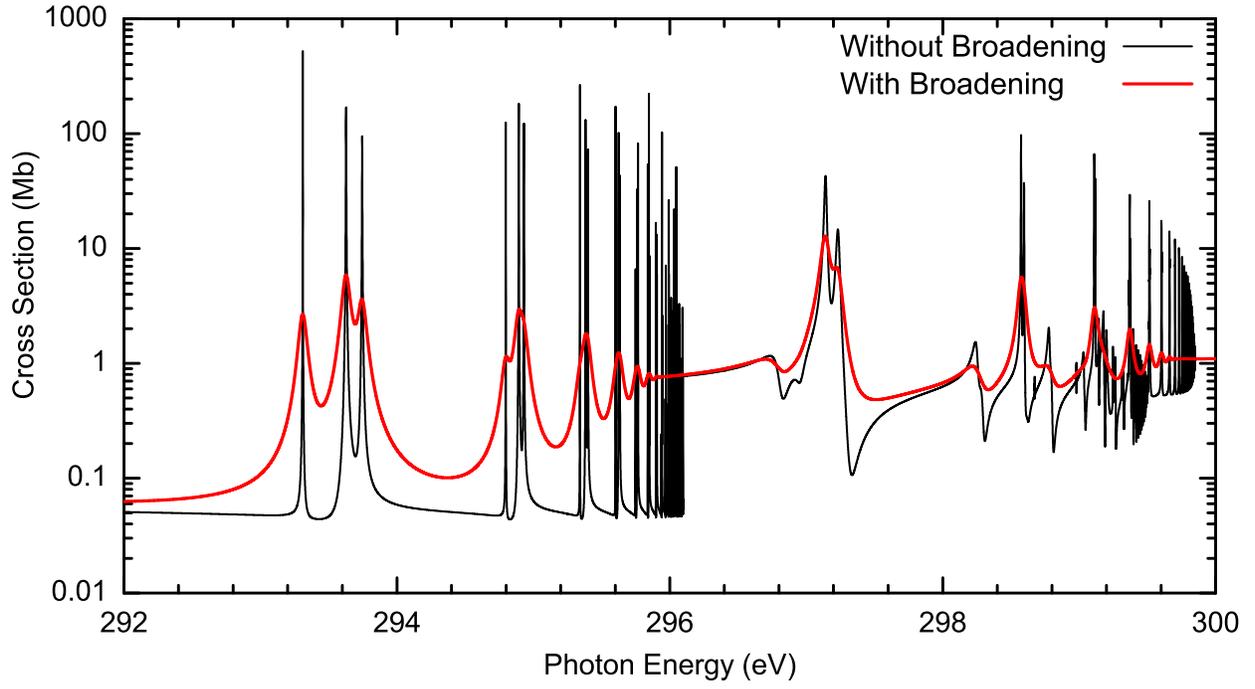}
\caption[Present R-matrix results with (red) and without (black) Auger broadening
effects for the ground state photoabsorption of \ion{C}{3} at the
K-edge.]{
\ion{C}{1} photoabsorption cross
sections for $n\ge 3$ without (black) and with (red) spectator Auger broadening
effects.}
\label{cpadamp}
\end{figure}

The importance of including spectator Auger decay channels via the optical potential approach can be seen in Fig.~\ref{cpadamp}.
Our results are shown with and without the spectator broadening effect.
Whereas the spectator-broadened results show the physically-correct constant
Auger width near threshold, the unbroadened results have widths that approach zero near threshold and are
therefore impossible to resolve using a finite set of energy mesh points.
Results are not shown for the lowest $1s2s^22p^2$ resonances
since these are not affected by spectator broadening.

\begin{figure}[!h]
\centering
\includegraphics[width=\textwidth, angle=0]{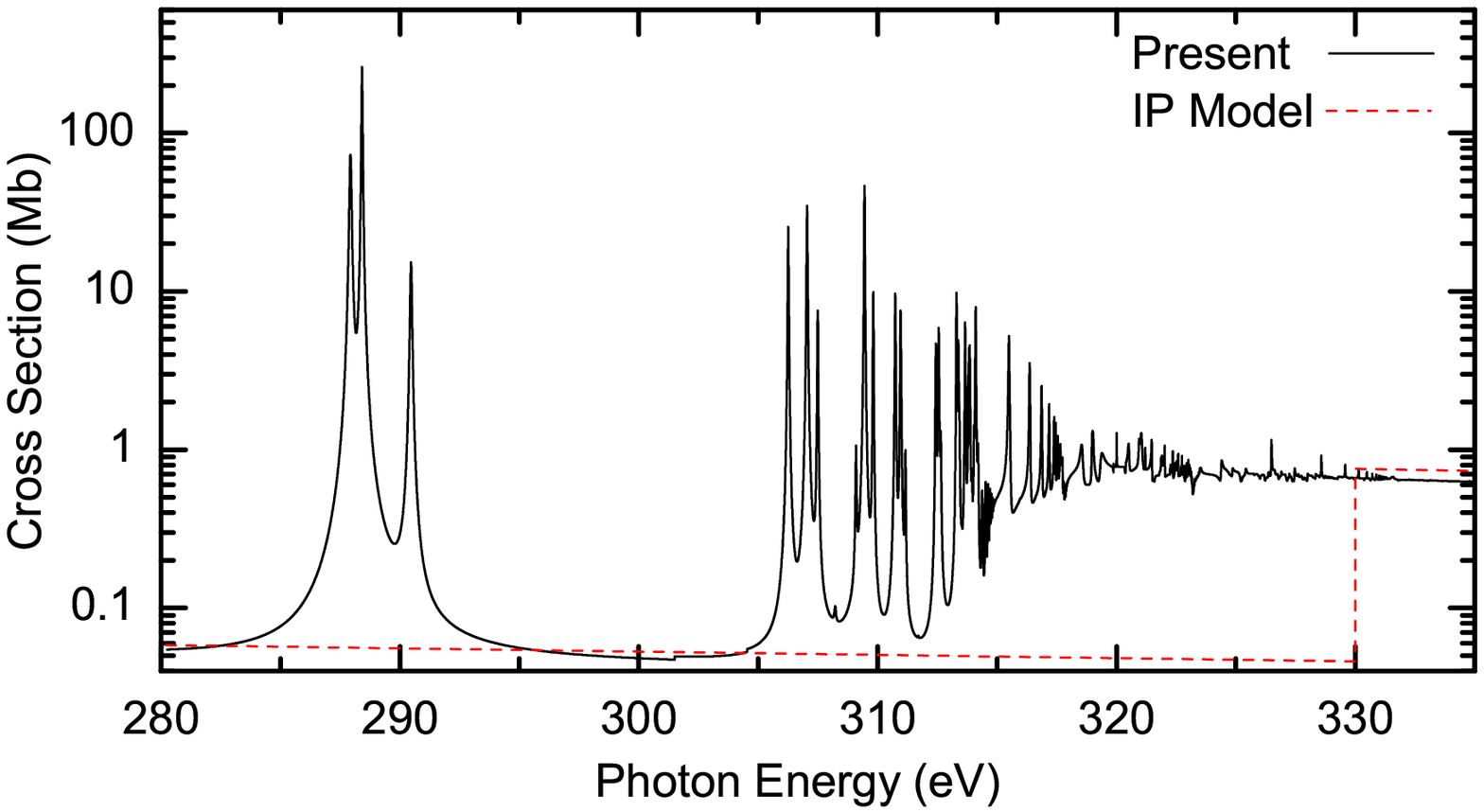}
\caption{\ion{C}{2} photoabsorption cross
sections.} \label{cIIpa}
\end{figure}

The K-shell cross section for \ion{C}{2} is shown in Fig.~\ref{cIIpa} for the entire $1s\rightarrow np$ photoabsorption resonance region,
and for the above-threshold $1s\rightarrow e^-$
photoionization region, where it is seen to be in fairly good agreement with the
IP results of \cite{reilman}.  We note that the IP results are tabulated on a sparse grid, and for \ion{C}{1} are provided at photon energies of
270 eV, 300 eV, and 330 eV, so that the results shown here are straight-line interpolations
and do not exhibit the unsmooth energy dependence found in our R-matrix cross sections.

\begin{figure}[!h]
\centering
\includegraphics[width=\textwidth]{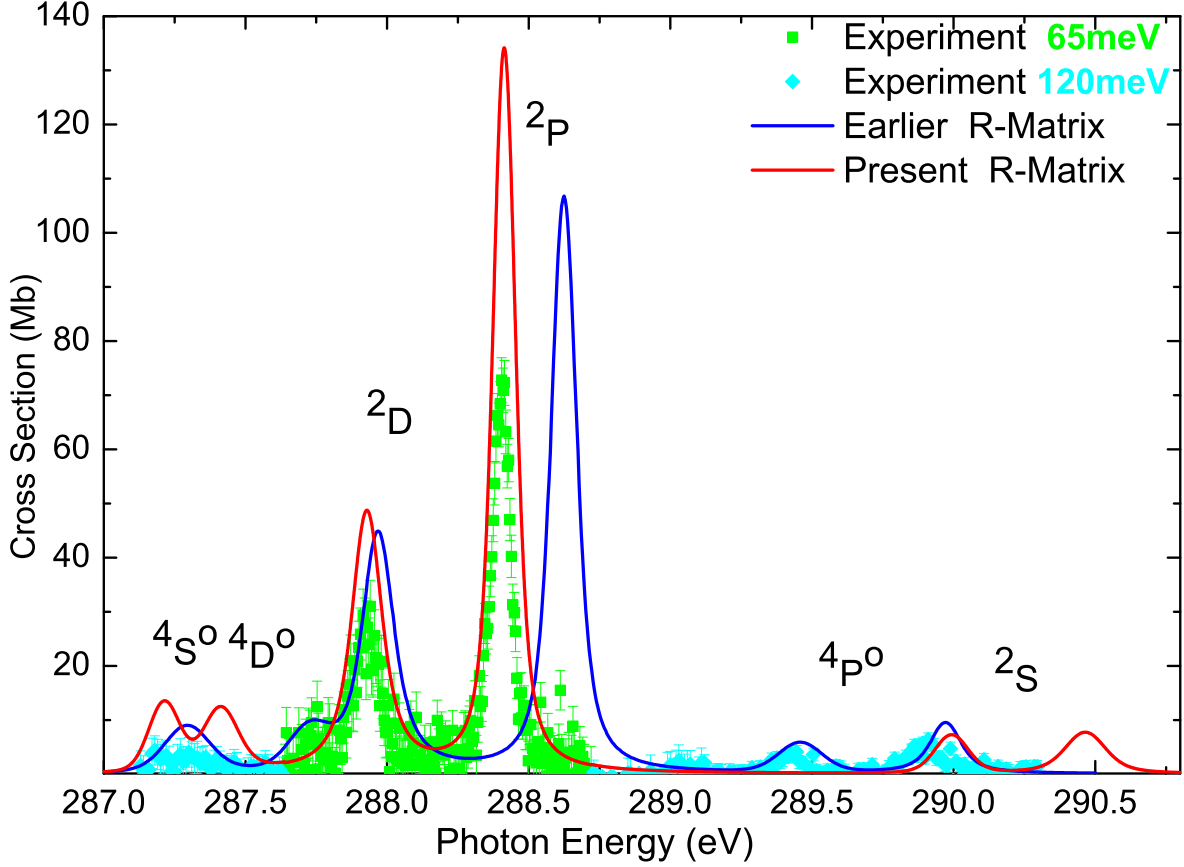}
\caption{The $1s \to
2p$ photoabsorption for an admixture of $80\%$ ground-state $1s^22s^22p(^2P)$ and
$20\%$ metastable state $1s^22s2p^2(^4P)$ \ion{C}{2} ions.
Experimental measurements~\citep{cIIexp} are
performed with two separate spectral resolutions of 65 meV and 120 meV. The theoretical cross sections are summed incoherently for an $80\%$ ground-state and  $20\%$ metastable state
admixture and then convoluted with a 65 meV or 120 meV FWHM Gaussian. Also shown are earlier R-matrix results~\citep{cIIexp}} \label{cIIexc}
\end{figure}

For \ion{C}{2}, there also exist experimental and earlier R-matrix results,
but only for the $1s\rightarrow 2p$ resonances \citep{cIIexp}.  In that experiment,
in addition to the \ion{C}{2} $1s^22s^22p(^2P)$ ground state, there was a $\approx 20$\% $1s^22s2p^2(^4P)$
metastable-state fraction, giving additional resonances not
included in our calculations for Fig~\ref{cIIpa}.
Thus we performed an additional R-matrix photoabsorption calculation from the metastable state,
including all quartet, rather than doublet, total symmetries, and then added the ground and
metastable cross sections together with weightings of 80\% and 20\%, respectively.
These results are shown in Fig.~\ref{cIIexc} along with the earlier results.
For the two strongest $^2P$ and $^2D$ resonances, where the experimental resolution was 65 meV, we
find good agreement with the experimental resonance positions, but both
present and earlier R-matrix resonance strengths are found to be significantly greater than the measured strengths.
The second set of experimental results was taken at 120 meV for the weaker resonances, and here
we find that our resonance positions are in error by as much as 0.5 eV and our strengths are again greater than experiment.

\begin{figure}[!h]
\centering
\includegraphics[width=\textwidth, angle=0]{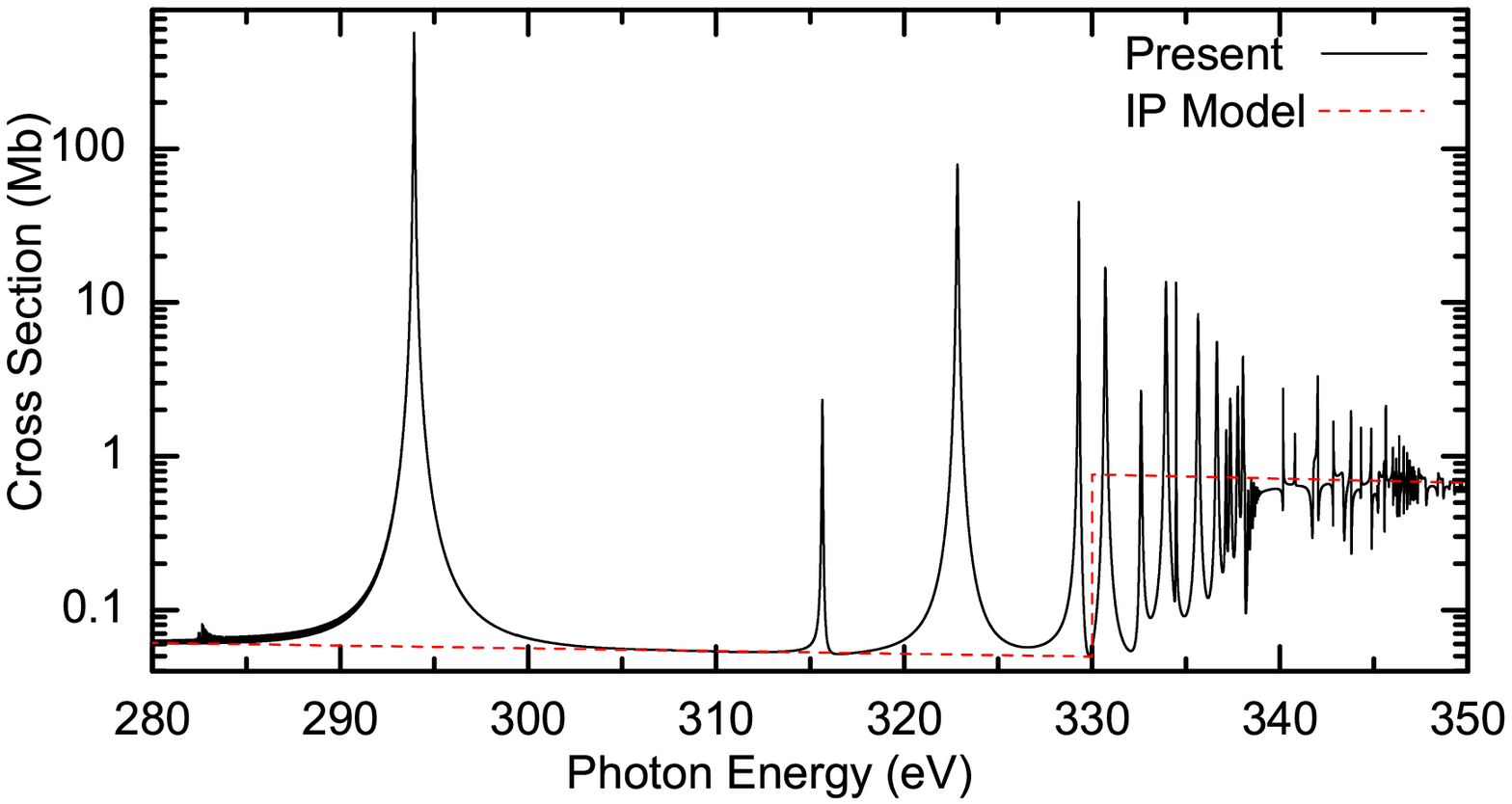}
\caption{\ion{C}{3} photoabsorption cross
sections.} \label{cIIIpa}
\end{figure}

\begin{figure}[!htb]
\centering
\includegraphics[width=4in]{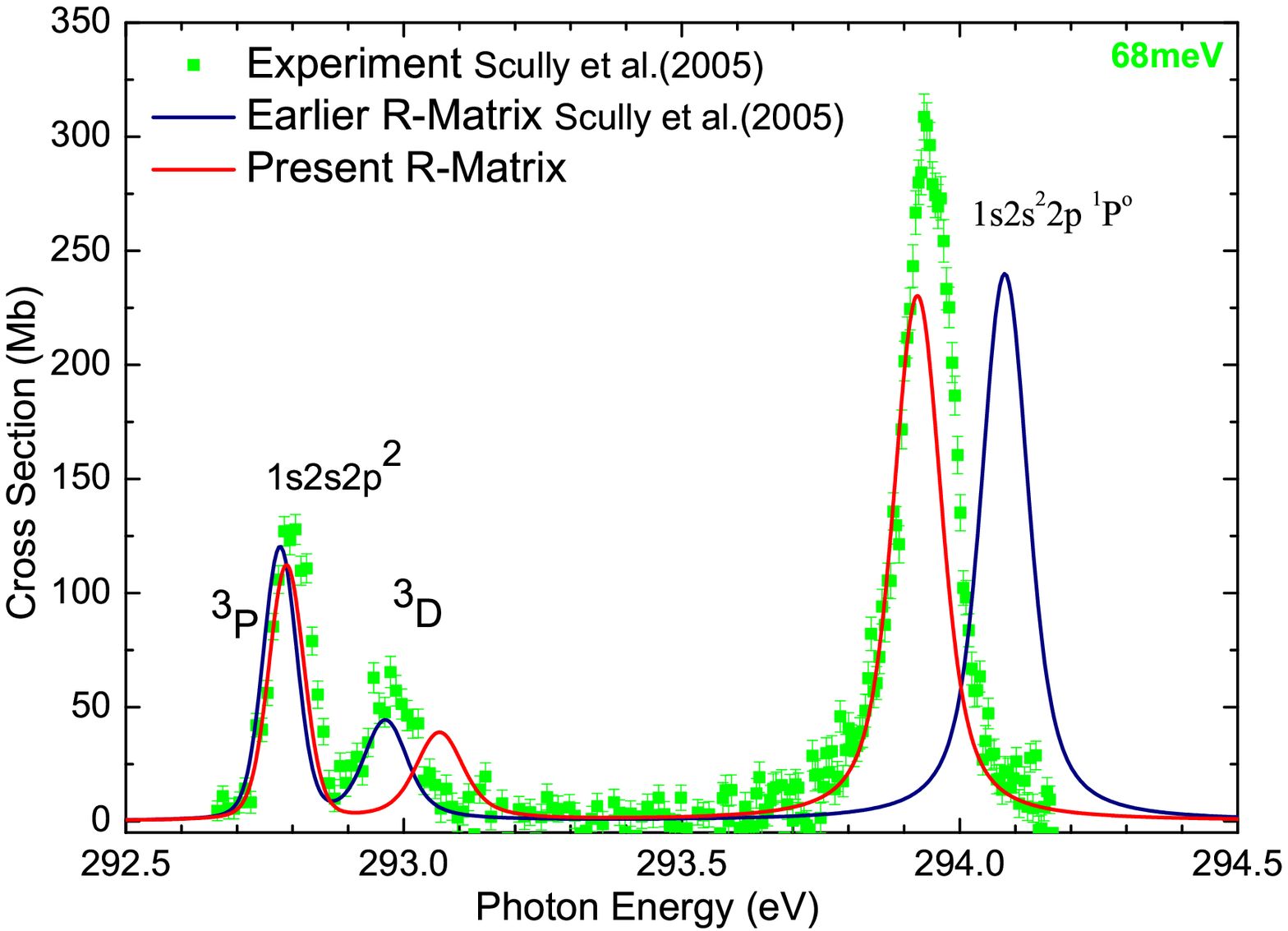}\\[0.05in]
\includegraphics[width=4.in]{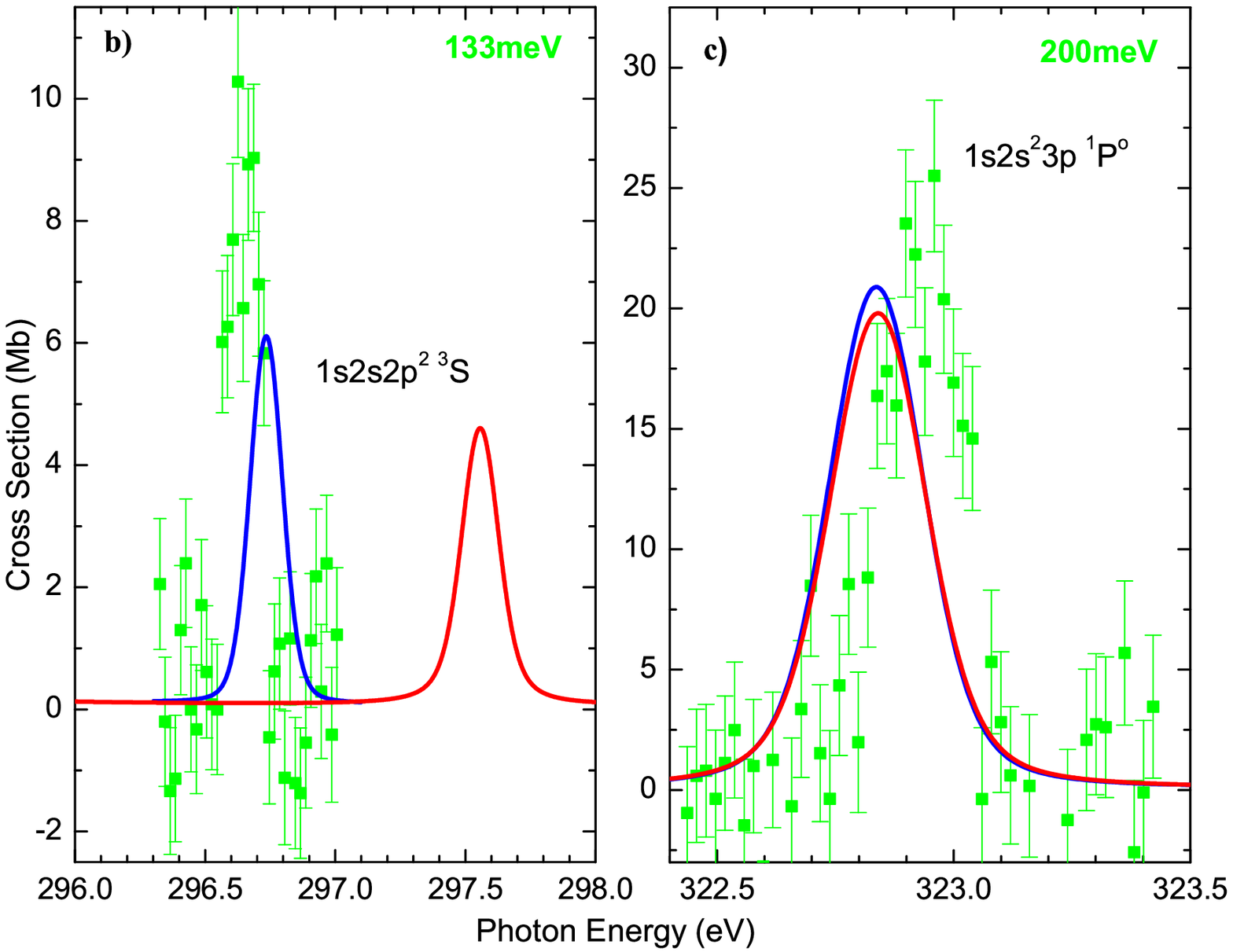}
\caption[Experimental measurements and present and earlier R-matrix
theoretical calculations of $1s \to2p {\rm\ and\ } 3p$ absorption
resonances of \ion{C}{3} ions.]
{Present R-matrix cross section vs. experimental
and earlier R-matrix results~\citep{cIIIexp}  for the $1s \to 2p,3p$ resonancs of \ion{C}{3}. The theoretical
curves are determined by considering an admixture of $68\%$ of the
ground state and $32\%$ of the metastable state and are convoluted with a FWHM Gaussian given by  the
experimental resolution. In \textbf{a)}, the
$1s2s2p^2\,(^3P,\, ^3D)$ and $1s2s^22p^2\,(^1P)$ resonances are shown at a spectral resolution of 68 meV, \textbf{b)} shows the
$1s2s2p^2\,(^3S)$ resonance at a resolution of 133 meV, and \textbf{c)}
shows the $1s2s^23p\,(^1P)$ resonance
at a spectral resolution of 200 meV.
}
\label{cIIIexc}
\end{figure}

The \ion{C}{3} photoabsorption cross section for the entire K-edge region is shown in Fig.~\ref{cIIIpa} and compared to the IP results above threshold, where
good agreement is again obtained. Results for the lowest $1s\rightarrow 2p,3p$
resonances are shown in Fig.~\ref{cIIIexc}.  For the dominant $1s2s^22p(^1P)$ resonance,
both R-matrix photoabsorption strengths are now less than the experimental value~\citep{cIIIexp},
but there is good agreement for the second-strongest $1s2s2p^2(^3P)$ resonance.

\begin{figure}[hbtp]
\centering
\includegraphics[width=\textwidth]{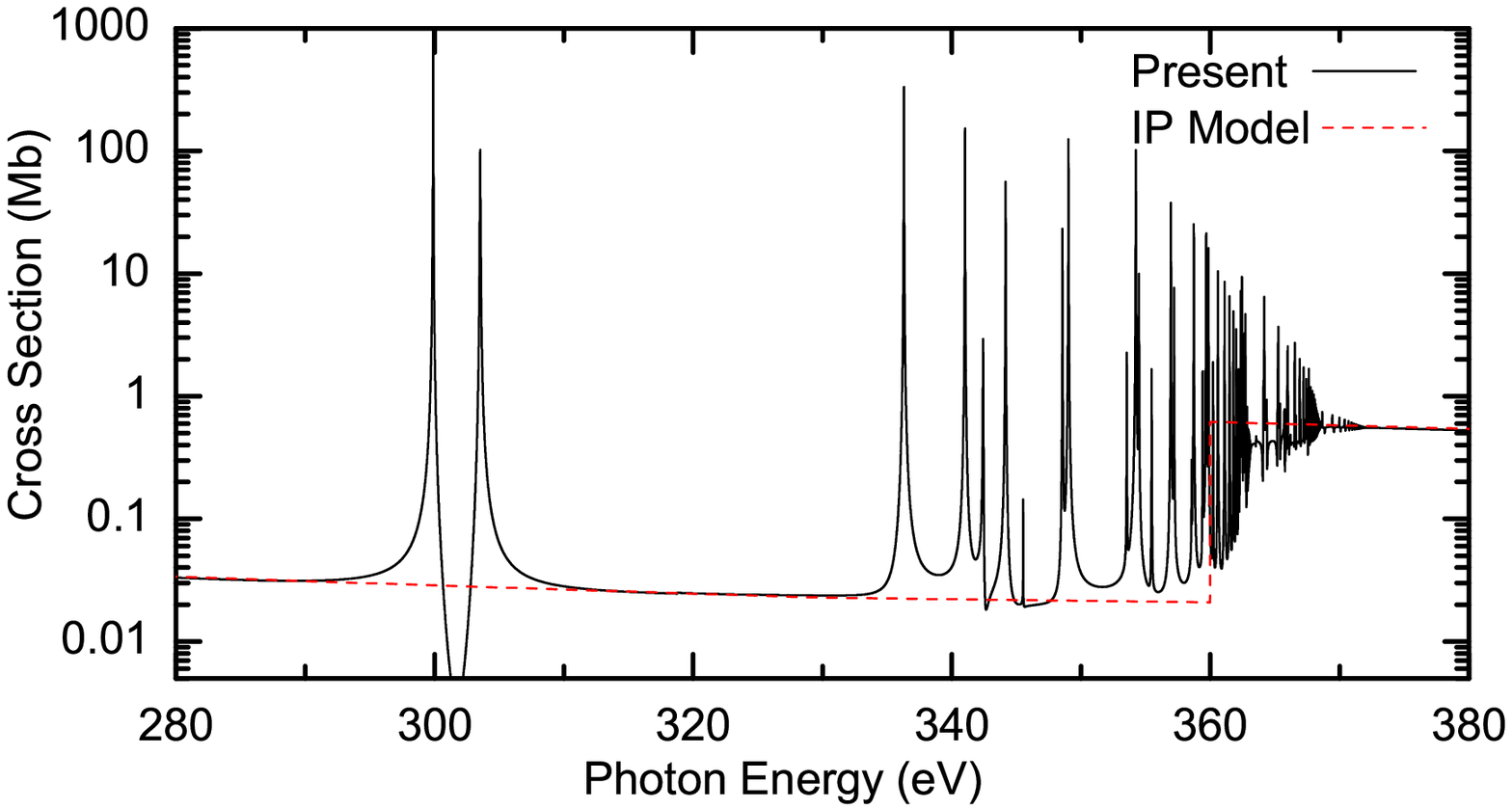}
\caption{\ion{C}{4} photoabsorption cross
sections.
} \label{cIVpa}
\end{figure}

The cross section for \ion{C}{4} is shown in Fig.~\ref{cIVpa}.
Here the resonances are given by $h\nu+1s^22s \rightarrow 1s2snp$, and there is no spectator channel.
Therefore, in order to be able to resolve the $n\rightarrow\infty$ narrowing resonances as threshold is approached,
we artificially broadened the entire series with a spectator width
of $\Gamma=0.027$ eV. This is less than the resolution  of the
ALS experimental measurements, which was given for \ion{C}{2} \citep{cIIexp}, \ion{C}{3}
\citep{cIIIexp}, and \ion{C}{4} \citep{cIVexp} as $\Delta E \ge 0.046$ eV, or the astrophysical measurement we describe,
which is given as $\Delta\lambda=0.023$~\AA, or  $\Delta E\approx 0.017$ eV for $E\approx 300$ eV.

\section{The C K-Edge in the X-Ray Spectrum of Mkn~421}

We have also compared our R-matrix photoabsorption cross sections with
the absorption features in the C~K-edge structure present in a {\it
Chandra} LETG+HRC-S spectrum of the blazar Mkn~421.  While the
LETG+HRC-S instrument has a strong C~K-edge feature arising from a
polyimide filter approximately 2750~\AA\ thick, there should be
additional absorption signatures present from the line-of-sight
absorption toward Mkn~421.

{\it Chandra} has observed Mkn~421 on several different occasions, but
one observation made by the LETG+HRC-S on 2003 July 1-2 (ObsID 4149)
caught the object in a particularly bright state and is of much higher
quality.  The observation and data are described in detail by \citet{nicastro2005}.

We retrieved standard pipeline-processed products from the {\it
Chandra} archive and constructed effective areas for the overlapping
spectral orders 1-10 using standard CIAO tasks.  Subsequent analysis
was performed using the Interactive Data Language-based Package for
Interactive Analysis of Line Emission \citep[PINTofALE;][]{kashyap2000}.
The adopted continuum model was optimized to the region around
the C~K-edge, ignoring the absorption edge structure.  A power law
continuum with photon index $\Gamma=2$ and interstellar medium (ISM)
absorption corresponding to a neutral hydrogen column density of
$N_H=1.5\times 10^{20}$~cm$^{-2}$ computed using the cross sections of \citet{balucinska1992} were adopted.  These values
are close (but not identical) to the values found by \citet{nicastro2005};
small differences can be ascribed to our optimization to the
C~K-edge region and to revisions in the instrument calibration between
their and our analyses.

\begin{figure}[h]
\centering
\includegraphics[height=\textwidth,angle=90]{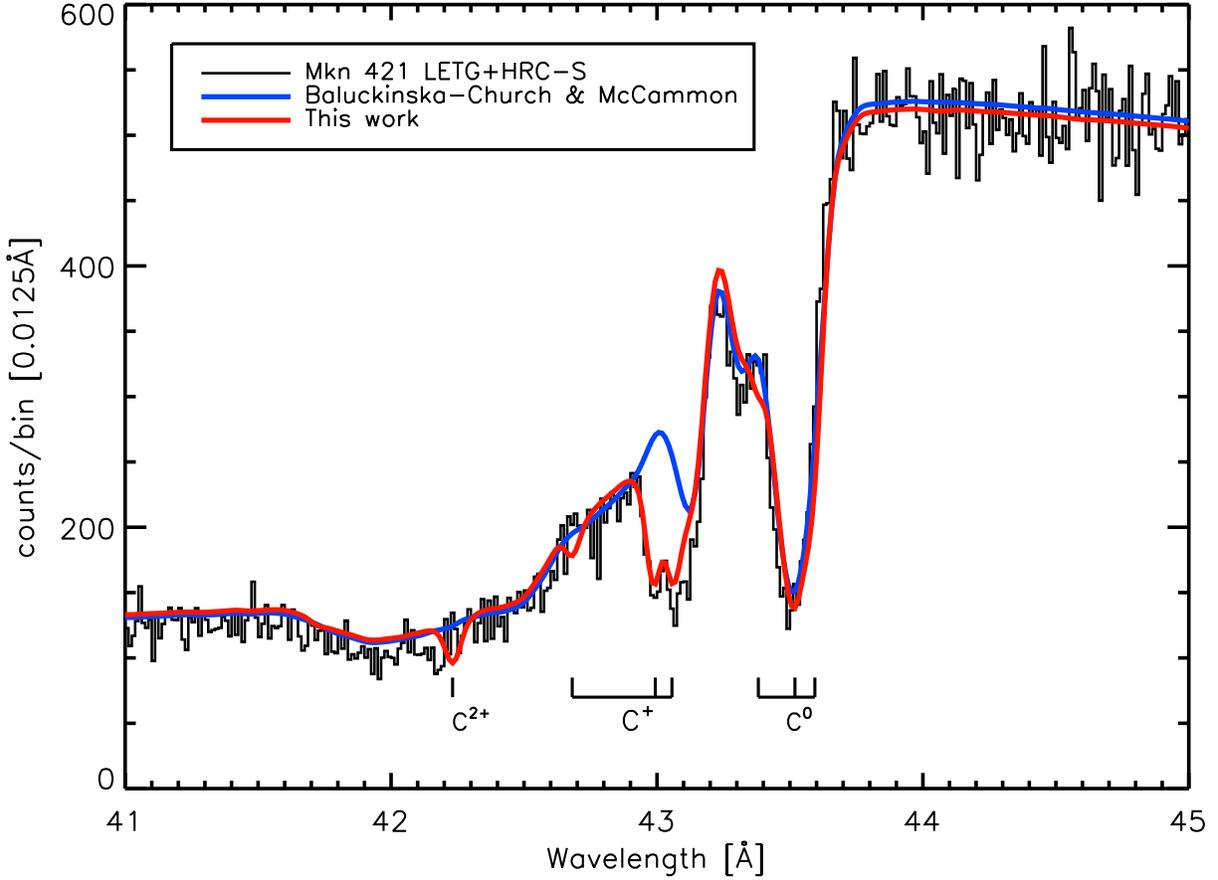}
\caption[Observed X-ray spectrum of the bright extra-galactic X-ray
source Blazar Mkn\ 421 near the carbon K-edge obtained from the
Chandra X-ray Observatory.]{\label{jdfig}The carbon K-edge region of the X-ray spectrum of the bright blazar
Mkn 421 observed by the {\it Chandra} LETG+HRC-S.  The edge absorption
is predominantly due to the polyimide UV-optical/ion blocking filter
on the HRC-S instrument, although ISM absorption contributions are
also present.  Two fits to a power-law continuum model with photon
index $\Gamma=2.0$, absorbed by an intervening ISM corresponding to a
neutral H column density of $1.5\times 1020$~cm$^{-2}$, are shown.
These differ significantly only in the carbon cross-sections employed:
the neutral \ion{C}{1} cross-section of \citet{balucinska1992};
and \ion{C}{1}, \ion{C}{2} and \ion{C}{3} cross-sections reported here.  In the latter
case, the C ion fractions were 20\%\ \ion{C}{1}, 60~\%\
\ion{C}{2} and 20\%\ \ion{C}{3}.  The effect of the \ion{C}{2} resonances are clearly
visible in the vicinity of 43~\AA. }
\end{figure}

While we could obtain a good match to the observations over the
continuum regions around the C~K-edge, this model systematically
underpredicted the data in the 42-44~\AA\ range by about 5-15\%.  We
ascribe this to uncertainty in the calibration---a notoriously
difficult task for X-ray instruments in the vicinity of the C~K-edge.
We applied a smooth broad Gaussian-like correction to the effective
area in order to ameliorate this effect.  The resulting model fit is
illustrated in Figure~\ref{jdfig}.  While the fit is generally very good, there
is a clear discrepancy near 43~\AA.
The~\citet{balucinska1992} cross section for carbon
is essentially that from the synthesis of \citet{henke1982,henke 1993}
combined with the C abundance of \citet{anders1982}, and amounts
to a simple step-function.  We computed a new ISM absorption
cross section replacing the neutral and ionized C cross sections with those from our R-Matrix
computations.  The division of carbon among different charge states
was adjusted by eye to obtain a good match to the data. This was
achieved with a mixture of 20\%\ \ion{C}{1}, 60\%\ \ion{C}{2}, and for illustrative
purposes, 20\%\ \ion{C}{3}.  The resulting model spectrum folded through
the instrument response is also illustrated in Figure~\ref{jdfig}.  There are
no obvious signs of significant \ion{C}{4} absorption in the data.

We do not place great weight on the C ion ratios used for the fit: the
instrument calibration would appear to require some revision before
quantitative measurements can be made.  However, we especially note
two aspects of the results. Firstly, the \ion{C}{2} resonance structure now
provides a good match to the observations in the vicinity of 43~\AA\
and we consider this a reliable detection of this species.  To our
knowledge, this represents the first X-ray identification of \ion{C}{2}
absorption in interstellar gas.  Secondly, there is a weak absorption
feature near 42.15~\AA\ in the observed spectrum that is reasonably
close to the \ion{C}{3} $1s2s^22p^2\,(^1P)$ resonance predicted by our
R-Matrix calculations.  The offset between the two is $\sim
0.05$~\AA\, or 0.35~eV.  Based on the comparison of the R-matrix
resonance energy and synchrotron observations illustrated in
Figure~\ref{jdfig}, an offset of 0.35~eV is much too large to be associated
with uncertainties in the predicted resonance position.  Instead, it
is possible that HRC-S imaging non-linearities could give rise to such
a discrepancy.  Line positions are generally expected to be better
than 0.01-0.02~\AA, but can occasionally be as far as 0.05~\AA\ out of
place.\footnote{E.g.\ {\it Chandra Proposer's Observatory Guide:
http://cxc.harvard.edu/proposer/POG}.}   We stop short of identifying
this absorption feature but draw attention to its possible interest
for future study.

\section{Summary and Conclusion}
Photoabsorption features of carbon ions (\ion{C}{1}-IV)) are studied using the R-matrix method, yielding, detailed information on the
carbon K-shell photoabsorption cross section spectra. Furthermore, we computed photoabsorption cross sections for the
additional \ion{C}{2}, \ion{C}{3}, and \ion{C}{4} isonuclear members, for which
synchrotron-facility measurements at the ALS and theoretical studies
already studied the lowest $1s \to  2p,3p$ resonance transitions.
Our computed resonance positions were within about 0.5 eV of the measured values,
and our strengths showed good agreement for some of the resonances
but were significantly too large or small, compared to experiment, for certain strong resonances.
These computed data are of particular importance for absorption studies of cosmic gas.
In turn, a more accurate description of the interstellar absorption near the C K-edge
in cosmic sources used as in-flight calibration standards should lead to refinements
in the calibration of spectrometers such as the {\it Chandra} LETGS.
Analysis of the LETGS spectrum of Mkn 421 has allowed us to identify interstellar absorption
due to \ion{C}{2} and estimate ion fractions of \ion{C}{1} and \ion{C}{2} for the first time using X-rays.

\section{Acknowledgment}
MFH, ShA, and TWG were supported in part by NASA APRA, NASA SHP SR\&T, and Chandra Project grants.
JJD was supported by NASA contract NAS8-39073 to the {\it Chandra} X-ray Center. BMMcL acknowledges support by the US
National Science Foundation through a grant to ITAMP at the Harvard-Smithsonian Center for Astrophysics.

\clearpage




\clearpage

\clearpage

\clearpage

\clearpage

\clearpage

\clearpage

\clearpage
\begin{figure}[!htb]
\centering
\includegraphics[width=\textwidth]{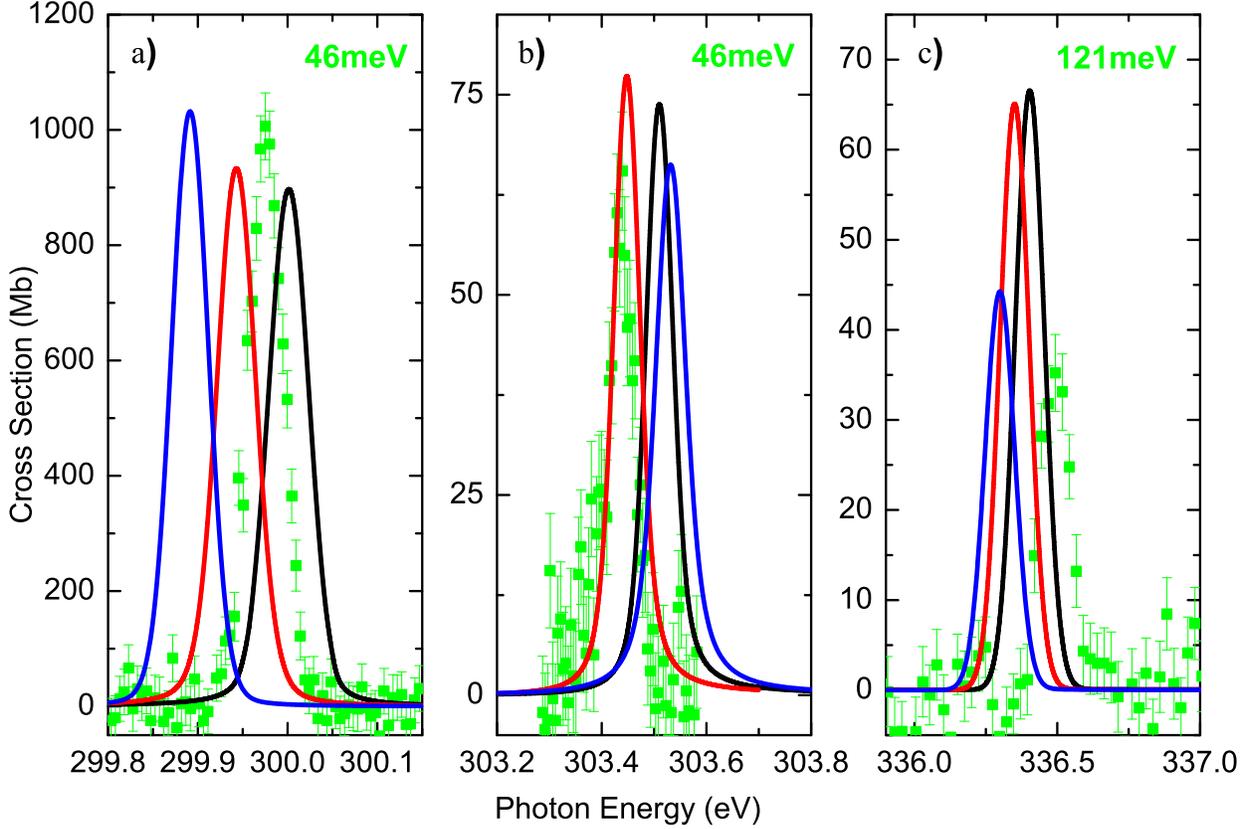}
\caption[Experimental measurements and present and earlier R-matrix
theoretical calculations of $1s \to2p {\rm\ and\ } 3p$ absorption
resonances of \ion{C}{4} ions.]{
Present R-matrix cross section vs. experimental
and earlier R-matrix results~\citep{cIVexp}  for the $1s \to 2p,3p$ resonancs of \ion{C}{4}. The green data points represent the experimental data, the red and black curves show previous R-matrix results in IC- and LS-coupling, respectively, and the blue curve shows the present R-matrix results
(all theoretical
cross sections are convoluted with a FWHM Gaussian given by  the
experimental resolution). In \textbf{a)} and \textbf{b)}, the $[1s (2s 2p)^3P]^2P$ and $[1s (2s 2p)^1P]^2P$ resonances, respectively, are measured with a spectral resolution of 46 meV, whereas for the $[1s (2s 3p)^3P]^2P$ resonance in \textbf{c)} , the
resolution is 121 meV.}
\label{cIVexc}
\end{figure}

\clearpage

\clearpage
\begin{table*}[!hbtp]
\caption{\label{cenergy} Energies (in Rydbergs) of the \ion{C}{2} target states and the \ion{C}{1} ground state. The $1s2s^22p^2\,(^4P)$ energy of 20.9133 Ryd relative to the $1s^22s^22p\,(^2P)$ ground state of \ion{C}{2} corresponds to a K-shell threshold energy of
296.098 eV relative to the \ion{C}{1} ground state, which compares favorably to the experimental value of $296.07\pm 0.2$ eV \cite{cexp}.
}\vspace*{0.07in}
\resizebox{3.in}{!}{
\begin{tabular}{l r r}  
\hline \hline
State          & R-matrix & NIST \\
\hline
$1s^22s^22p^2\,(^3P)$      & $-0.8493$   & $-0.8277     $ \\[-0.09in]
\hline
$1s^22s^22p\,(^2P)$        & $0.0000 $   & $ 0.0000     $ \\[-0.09in]
$1s^22s2p^2\,(^4P)$        & $0.3814 $   & $ 0.3922     $ \\[-0.09in]
$1s^22s2p^2\,(^2D)$        & $0.7128 $   & $ 0.6828     $ \\[-0.09in]
$1s^22s2p^2\,(^2S)$        & $0.9850 $   & $ 0.8793     $ \\[-0.09in]
$1s^22s2p^2\,(^2P)$        & $1.0450 $   & $ 1.0083     $ \\[-0.09in]
$1s^22p^3\,(^4S)$          & $1.3104 $   & $ 1.2942     $ \\[-0.09in]
$1s^22p^3\,(^2D)$          & $1.4311 $   & $ 1.3711     $ \\[-0.09in]
$1s^22p^3\,(^2P)$          & $1.6792 $   & $ 1.5377     $ \\[-0.09in]
\hline
$1s2s^22p^2\,(^4P)$        & $20.9133$   &  \\[-0.09in]
$1s2s^22p^2\,(^2D)$        & $21.1651$   &                        \\[-0.09in]
$1s2s^22p^2\,(^2P)$        & $21.1899$   &                        \\[-0.09in]
$1s2s^22p^2\,(^2S)$        & $21.3351$   &                        \\[-0.09in]
$1s2s(^1S)2p^3\,(^4S)$     & $21.4753$   &                        \\[-0.09in]
$1s2s(^3S)2p^3\,(^4D)$     & $21.5097$   &                        \\[-0.09in]
$1s2s(^3S)2p^3\,(^4P)$     & $21.6968$   &                        \\[-0.09in]
$1s2s(^1S)2p^3\,(^2D)$     & $21.8446$   &                        \\[-0.09in]
$1s2s(^1S)2p^3\,(^2P)$     & $22.0355$   &                        \\[-0.09in]
$1s2s(^3S)2p^3\,(^4S)$     & $22.0847$   &                        \\[-0.09in]
$1s2s(^3S)2p^3\,(^2D)$     & $22.1103$   &                        \\[-0.09in]
$1s2s(^3S)2p^3\,(^2S)$     & $22.2898$   &                        \\[-0.09in]
$1s2s(^3S)2p^3\,(^2P)$     & $22.3180$   &                        \\[-0.09in]
$1s2p^4\,(^4P)$            & $22.4901$   &                        \\[-0.09in]
$1s2p^4\,(^2D)$            & $22.6795$   &                        \\[-0.09in]
$1s2p^4\,(^2P)$            & $22.7300$   &                        \\[-0.09in]
$1s2p^4\,(^2S)$            & $23.0860$   &                        \\[-0.05in]
\hline \hline
\end{tabular}
}
\end{table*}

\clearpage
\begin{table*}[!hbtp]

\caption{\label{cIIgsenergy}
Energies (in Rydbergs) of the \ion{C}{3} target states and the \ion{C}{2} ground state. } \vspace*{0.07in}
\resizebox{3in.}{!}{
\begin{tabular}{l r r}  
\hline \hline
State                 & R-matrix & NIST \\
\hline
$1s^22s^22p\,(^2P)$   & $-1.7977$ & $-1.7921 $ \\[-0.09in]
\hline
$1s^22s^2\,(^1S)$     & $0.0000 $ & $ 0.0000 $ \\[-0.09in]
$1s^22s2p\,(^3P)$     & $0.4745 $ & $ 0.4777 $ \\[-0.09in]
$1s^22s2p\,(^1P)$     & $0.9570 $ & $ 0.9327 $ \\[-0.09in]
$1s^22p^2\,(^3P)$     & $1.2649 $ & $ 1.2528 $ \\[-0.09in]
$1s^22p^2\,(^1D)$     & $1.3630 $ & $ 1.3293 $ \\[-0.09in]
$1s^22p^2\,(^1S)$     & $1.7881 $ & $ 1.6632 $ \\[-0.09in]
\hline
$1s2s^22p\,(^3P)$     & $21.3622$ &            \\[-0.09in]
$1s2s^22p\,(^1P)$     & $21.5849$ &            \\[-0.09in]
$1s2s(^1S)2p^2\,(^3P)$& $21.9689$ &            \\[-0.09in]
$1s2s(^3S)2p^2\,(^3D)$& $21.9997$ &            \\[-0.09in]
$1s2s(^3S)2p^2\,(^1D)$& $22.3201$ &            \\[-0.09in]
$1s2s(^3S)2p^2\,(^3S)$& $22.3269$ &            \\[-0.09in]
$1s2s(^3S)2p^2\,(^3P)$& $22.4088$ &            \\[-0.09in]
$1s2s(^3S)2p^2\,(^1P)$& $22.5674$ &            \\[-0.09in]
$1s2s(^1S)2p^2\,(^1S)$& $22.6511$ &            \\[-0.09in]
$1s2p^3\,(^3D)$       & $22.6802$ &            \\[-0.09in]
$1s2p^3\,(^3S)$       & $22.7780$ &            \\[-0.09in]
$1s2p^3\,(^1D)$       & $22.8708$ &            \\[-0.09in]
$1s2p^3\,(^3P)$       & $22.9926$ &            \\[-0.09in]
$1s2p^3\,(^1P)$       & $23.1865$ &
\\*[0.1in] \hline \hline
\end{tabular}
}
\end{table*}

\clearpage
\begin{table*}[!h]
\label{cIIIgsenergy}
\caption{
Energies (in Rydbergs) of the \ion{C}{4} target states and the \ion{C}{3} ground state.
} \vspace*{0.07in}
\resizebox{3.in}{!}{
\begin{tabular}{l r r}  
\hline \hline
State               & R-matrix & NIST\\
\hline
$1s^22s^2\,(^1S)$     & $-3.5264$ & $-3.5197$   \\[-0.09in]
\hline
$1s^22s\,(^2S)$       & $0.0000 $ & $0.0000 $   \\[-0.09in]
$1s^22p\,(^2P)$       & $0.5903 $ & $0.5883 $   \\[-0.09in]
\hline
$1s2s^2\,(^2S)$       & $21.3887$ & $       $   \\[-0.09in]
$1s2s(^1S)2p \,(^2P)$ & $22.0009$ & $       $   \\[-0.09in]
$1s2s(^3S)2p \,(^2P)$ & $22.2608$ & $       $   \\[-0.09in]
$1s2p^2\,(^2D)$       & $22.4955$ & $       $   \\[-0.09in]
$1s2p^2\,(^2P)$       & $22.5767$ & $       $   \\[-0.09in]
$1s2p^2\,(^2S)$       & $23.0266$ & $       $   \\[-0.00in]
\hline \hline
\end{tabular}
}
\end{table*}

\clearpage
\begin{table*}[!hbtp]
\caption{\label{cIVenergy}
Energies (in Rydbergs) of the \ion{C}{5} target states and the \ion{C}{4} ground state.
\vspace*{0.07in}}
\resizebox{3.in}{!}{
\begin{tabular}{l r r}  
\hline \hline
State               & R-matrix & NIST\\
\hline
$1s^22s\,(^2S)       $  & $-4.7402$ & -4.7402  \\[-0.09in]
\hline
$1s^2\,(^1S)         $  & $ 0.0000$ &  0.0000  \\[-0.09in]
$1s2s\,(^3S)         $  & $21.9670$ & 21.9731  \\[-0.09in]
$1s2s\,(^1S)         $  & $22.3662$ & 22.3718  \\[-0.09in]
$1s2p\,(^3P)         $  & $22.3673$ & 22.3737  \\[-0.09in]
$1s2p\,(^1P)         $  & $22.6236$ & 22.6302  \\*[0.00in]
\hline \hline
\end{tabular}
}
\end{table*}

\clearpage
\begin{table*}[!h]
\caption{\label{cauger}
R-matrix Auger widths (in eV) for the 17 \ion{C}{2} autoionizing
target states above the K-shell threshold (see Table~\ref{cenergy}).
Also shown are level-averaged MBCP \citep{km3} and level-averaged MCDF \citep{chenb} widths.}
\vspace*{0.07in} \centering
\begin{tabular}{r l r r r r c r}  
\hline \hline
 & State  &        & R-matrix & & MCBP & & MCDF \\
\cline{2-2}\cline{4-4}\cline{6-6}\cline{8-8}
1 & $1s2s^22p^2\,(^4P)$    & & $6.45E-02$ & & $8.61E-02$ & & $6.68E-02$\\[-0.09in]
2 & $1s2s^22p^2\,(^2D)$    & & $9.14E-02$ & & $1.16E-01$ & & $8.62E-02$\\[-0.09in]
3 & $1s2s^22p^2\,(^2P)$    & & $4.93E-02$ & & $5.19E-02$ & & $4.75E-02$\\[-0.09in]
4 & $1s2s^22p^2\,(^2S)$    & & $8.65E-02$ & & $1.02E-01$ & & $1.49E-04$\\[-0.09in]
5 & $1s2s(^1S)2p^3\,(^4S)$ & & $1.55E-02$ & & $2.06E-02$ & & $5.86E-02$\\[-0.09in]
6 & $1s2s(^3S)2p^3\,(^4D)$ & & $4.60E-02$ & & $6.42E-02$ & & $4.77E-02$\\[-0.09in]
7 & $1s2s(^3S)2p^3\,(^4P)$ & & $3.56E-02$ & & $4.87E-02$ & & $3.73E-02$\\[-0.09in]
8 & $1s2s(^1S)2p^3\,(^2D)$ & & $7.21E-02$ & & $9.64E-02$ & & $9.08E-02$\\[-0.09in]
9 & $1s2s(^1S)2p^3\,(^2P)$ & & $6.42E-02$ & & $7.77E-02$ & & $8.01E-02$\\[-0.09in]
10& $1s2s(^3S)2p^3\,(^4S)$ & & $4.57E-02$ & & $7.15E-02$ & & $2.49E-02$\\[-0.09in]
11& $1s2s(^3S)2p^3\,(^2D)$ & & $7.78E-02$ & & $1.18E-01$ & & $8.23E-02$\\[-0.09in]
12& $1s2s(^3S)2p^3\,(^2S)$ & & $1.49E-02$ & & $1.25E-02$ & & $8.89E-03$\\[-0.09in]
13& $1s2s(^3S)2p^3\,(^2P)$ & & $6.56E-02$ & & $1.01E-01$ & & $7.13E-02$\\[-0.09in]
14& $1s2p^4\,(^4P)$        & & $3.89E-03$ & & $6.83E-02$ & & $5.18E-02$\\[-0.09in]
15& $1s2p^4\,(^2D)$        & & $7.24E-02$ & & $1.08E-01$ & & $8.16E-02$\\[-0.09in]
16& $1s2p^4\,(^2P)$        & & $4.60E-02$ & & $6.69E-02$ & & $5.03E-02$\\[-0.09in]
17& $1s2p^4\,(^2S)$        & & $4.90E-02$ & & $1.36E-01$ & & $5.56E-02$\\[-0.02in]
\hline \hline
\\[-.4in]
\end{tabular}
\end{table*}

\clearpage
\begin{table*}[!h]
\begin{minipage}[t]{\textwidth}
\caption{\label{cIIauger}
R-matrix Auger widths (in eV) for the \ion{C}{3} autoionizing
target states above the K-shell threshold (see Table~\ref{cIIgsenergy}).
Also shown are level-averaged MBCP \citep{km1} and level-averaged MCDF \citep{chenb} widths. \vspace*{0.07in}} \centering
\resizebox{5.in}{!}{
\begin{tabular}{r l r r r r c r}  
\hline \hline
 & State  &        & R-matrix$^{{\rm a}}$ & & MCBP & & MCDF \\
\cline{2-2}\cline{4-4}\cline{6-6}\cline{8-8}
1  & $1s2s^22p\,(^3P)$      & & $ 7.18E-02 $ & & $ 7.93E-02 $ & & $ 6.72E-02 $\\[-0.09in]
2  & $1s2s^22p\,(^1P)$      & & $ 5.46E-02 $ & & $ 5.30E-02 $ & & $ 4.76E-02 $\\[-0.09in]
3  & $1s2s(^1S)2p^2\,(^3P)$ & & $ 1.10E-02 $ & & $ 1.38E-02 $ & & $ 2.47E-02 $\\[-0.09in]
4  & $1s2s(^3S)2p^2\,(^3D)$ & & $ 4.76E-02 $ & & $ 5.17E-02 $ & & $ 4.29E-02 $\\[-0.09in]
5  & $1s2s(^3S)2p^2\,(^1D)$ & & $ 9.07E-02 $ & & $ 1.14E-01 $ & & $ 1.13E-01 $\\[-0.09in]
6  & $1s2s(^3S)2p^2\,(^3S)$ & & $ 2.61E-02 $ & & $ 2.39E-02 $ & & $ 2.17E-02 $\\[-0.09in]
7  & $1s2s(^3S)2p^2\,(^3P)$ & & $ 4.64E-02 $ & & $ 5.92E-02 $ & & $ 4.91E-02 $\\[-0.09in]
8  & $1s2s(^3S)2p^2\,(^1P)$ & & $ 1.70E-02 $ & & $ 1.17E-02 $ & & $ 7.96E-03 $\\[-0.09in]
9  & $1s2s(^1S)2p^2\,(^1S)$ & & $ 7.46E-02 $ & & $ 7.98E-02 $ & & $ 8.36E-02 $\\[-0.09in]
10 & $1s2p^3\,(^3D)$        & & $ 5.76E-02 $ & & $ 7.15E-02 $ & & $ 6.08E-02 $\\[-0.09in]
11 & $1s2p^3\,(^3S)$        & & $   -      $ & & $   -      $ & & $ 1.17E-06 $\\[-0.09in]
12 & $1s2p^3\,(^1D)$        & & $ 5.99E-02 $ & & $ 7.19E-02 $ & & $ 6.02E-02 $\\[-0.09in]
13 & $1s2p^3\,(^3P)$        & & $ 3.56E-02 $ & & $ 4.19E-02 $ & & $ 3.67E-02 $\\[-0.09in]
14 & $1s2p^3\,(^1P)$        & & $ 3.60E-02 $ & & $ 3.90E-02 $ & & $ 3.53E-02 $\\[0.00in]
\hline \hline
\end{tabular}
}
\end{minipage}
\\[0.1in]
\end{table*}

\begin{table*}[!h]
\caption{\label{cIIIauger}
R-matrix Auger widths (in eV) for the \ion{C}{4} autoionizing
target states above the K-shell threshold (see Table~\ref{cIIIgsenergy}).
Also shown are level-averaged MBCP \citep{km2} and level-averaged MCDF \citep{chenb} widths.
}\vspace*{0.07in}
\begin{minipage}[t]{\textwidth}
\centering
\resizebox{3.5in}{!}{
\begin{tabular}{r l r r r r}  
\hline \hline
 & State  &        & R-matrix & & MCBP \\
\cline{2-2}\cline{4-4}\cline{6-6}
1  & $1s2s^2\,(^2S)$       & & $ 6.99E-02 $ & & $ 7.12E-02 $\\[-0.09in]
2  & $1s2s(^1S)2p \,(^2P)$ & & $ 3.95E-03 $ & & $          $\\[-0.09in]
3  & $1s2s(^3S)2p \,(^2P)$ & & $ 3.64E-02 $ & & $          $\\[-0.09in]
4  & $1s2p^2\,(^2D)$       & & $ 5.48E-02 $ & & $ 6.07E-02 $\\[-0.09in]
5  & $1s2p^2\,(^2P)$       & & $          $ & & $          $\\[-0.09in]
6  & $1s2p^2\,(^2S)$       & & $ 8.76E-03 $ & & $ 5.22E-03 $\\[-0.00in]
\hline \hline
\end{tabular}
}
\end{minipage}
\end{table*}

\end{document}